\documentclass[12pt]{article}

\def\fnote#1#2{\begingroup\def\thefootnote{#1}\footnote{#2}\addtocounter{footnote}{-1}\endgroup}

\usepackage{amssymb}
\usepackage{makeidx}
\usepackage{epsf}
\usepackage{graphicx}        

\makeindex

\def\inbar{\vrule height1.5ex width.4pt depth0pt}
\def\IB{\relax{\rm I\kern-.18em B}}
\def\IC{\relax\,\hbox{$\inbar\kern-.3em{\rm C}$}}
\def\ID{\relax{\rm I\kern-.18em D}}
\def\IE{\relax{\rm I\kern-.18em E}}
\def\IF{\relax{\rm I\kern-.18em F}}
\def\IG{\relax\,\hbox{$\inbar\kern-.3em{\rm G}$}}
\def\IH{\relax{\rm I\kern-.18em H}}
\def\II{\relax{\rm I\kern-.18em I}}
\def\IK{\relax{\rm I\kern-.18em K}}
\def\IL{\relax{\rm I\kern-.18em L}}
\def\IM{\relax{\rm I\kern-.18em M}}
\def\IN{\relax{\rm I\kern-.18em N}}
\def\IO{\relax\,\hbox{$\inbar\kern-.3em{\rm O}$}}
\def\IP{\relax{\rm I\kern-.18em P}}
\def\IQ{\relax\,\hbox{$\inbar\kern-.3em{\rm Q}$}}
\def\IR{\relax{\rm I\kern-.18em R}}
\def\IT{\relax{\rm I\kern-.18em T}}
\def\ZZ{\relax{\sf Z\kern-.4em Z}}

\def\nablaslash{\relax{\rm /\kern-.28em \nabla}}

       \def\g{\gamma}  
\def\e{\epsilon} \def\G{\Gamma}   \def\k{\kappa}  \def\l{\lambda}
  \def\om{\omega}   \def\si{\sigma}

\def\cA{{\cal A}} 
\def\cC{{\cal C}}   \def\cF{{\cal F}}
 \def\cH{{\cal H}}

\def\afrak{{\mathfrak a}}    
\def\gfrak{{\mathfrak g}}    \def\kfrak{{\mathfrak k}}
\def\lfrak{{\mathfrak l}}     
\def\nfrak{{\mathfrak n}}   \def\pfrak{{\mathfrak p}}
\def\sfrak{{\mathfrak s}}

\def\mathC{{\mathbb C}}   
   
\def\mathR{{\mathbb R}}     \def\mathZ{{\mathbb Z}}

 \def\oD3{{\overline \rmD 3}}

  \def\otau{{\overline \tau}}

   \def\Hdot{{\dot{H}}}  

\def\phidot{{\dot{\phi}}}

\def\fnote#1#2{\begingroup\def\thefootnote{#1}\footnote{#2}\addtocounter
{footnote}{-1}\endgroup}
\def\beq{\begin{equation}}
\def\eeq{\end{equation}}
\def\bea{\begin{eqnarray}}
\def\eea{\end{eqnarray}}

\def\lleq#1{\label{#1}\eeq}
\let\nn=\nonumber

\def\notin{\ \hbox{{$\in$}\kern-.51em\hbox{/}}}
\def\notsubset{\ \hbox{{$\subset$}\kern-.63em\hbox{/}}}
\def\lra{\longrightarrow}

\def\del{\partial}

  \def\E1Fq{E_1/\IF_q}

\def\rmD{{\rm D}}

      \def\rmGL{{\rm GL}}

\def\rmSL{{\rm SL}}

   \def\rmdet{{\rm det}}      \def\rmdim{{\rm dim}}

\def\rmrk{{\rm rk}}
   \def\rmmod{{\rm mod}}

 \def\rmth{{\rm th}}      \def\rmtr{{\rm tr}}

 \def\rmD{{\rm D}}

\def\rmad{{\rm ad}}

      \def\rmdet{{\rm det}}    
 \def\rmdim{{\rm dim}}

\def\rmmod{{\rm mod}}

 \def\rmrk{{\rm rk}}

        \def\rmtr{{\rm tr}}

           \def\rmGL{{\rm GL}}

      \def\rmIm{{\rm Im}}

        \def\rmPl{{\rm Pl}}

       \def\rmSL{{\rm SL}}
\def\rmSO{{\rm SO}}
\def\rmSp{{\rm Sp}}

\def\notdiv{{\relax{~|\kern-.35em /~}}}

\def\boxit#1{
\vbox{\hrule height1pt\hbox{\vrule width1pt\kern0.3cm
\vbox{\kern0.3cm\hbox{$\displaystyle#1$}\kern0.3cm}\kern0.3cm\vrule
width1pt}\hrule height1pt}}

\begin{document}
\parindent=0pt


\phantom{whatever}

\vskip 1.1truein
 \baselineskip=19pt

 \centerline{\large {\bf A General Framework of Automorphic Inflation }}


\vskip .4truein

\centerline{\sc Rolf Schimmrigk\fnote{$\diamond$}{
  netahu@yahoo.com; rschimmr@iusb.edu}}

\vskip .2truein

\centerline{Dept. of Physics, Indiana University at South Bend}
\centerline{1700 Mishawaka Ave. South Bend, IN 46634}

\vskip 1.4truein

\centerline{\bf Abstract.}
\parskip=0pt
 \begin{quote}
    Automorphic inflation is an application of the framework of
    automorphic scalar field theory, based on the theory of automorphic
    forms and representations. In this paper the general framework of
    automorphic and modular inflation is described in some detail, with
    emphasis on the resulting stratification of the space of scalar field
    theories in terms of the group theoretic data associated to the shift
    symmetry, as well as the automorphic data that specifies the potential.
    The class of theories based on Eisenstein series provides a natural
    generalization of the model of $j$-inflation considered previously.
 \end{quote}

\parskip=0.15truein

\renewcommand\thepage{}
\newpage
\pagenumbering{arabic}

\baselineskip=19pt
\parskip=0.05truein

\tableofcontents

\vfill \eject


\baselineskip=22.1pt
\parskip=0.15truein

\section{Introduction}

The space of interacting scalar fields is of interest in a variety
of problems that arise in quite different fields in physics.
 It has however
 remained a somewhat amorphous object without much structure.
 One familiar organizing tool is to consider field multiplets $\phi^I$,
 $I = 1,...,n$,
 that carry representations of some continuous group $G$. Discrete groups have played a less
 prominent role, but have recently been considered as a way to introduce structure into
 the set of such theories via a stratification of this space.
  Theories in this framework are based on the choice of a continuous group
  $G$ and  a discrete subgroup $\G \subset G(\mathZ)$ of the group $G$ considered over the
rational  integers $\mathZ$. Associated to such a pair of groups
$(G,\G)$  is a space  of automorphic forms $\Phi$, viewed as
 functions on $G$ with certain properties. These functions descend to a curved quotient space
 $X$ derived from $(G,\G)$ that defines the target space of the fields $\phi^I$.
This set-up leads to a highly structured field theory space
$\cF(G,\G, \Phi)$ that can be used as a starting point for models
characterized by potentials $V(\phi^I)$ that are derived from
automorphic forms on $G$, and a kinetic term determined by a
target space metric $G_{IJ}$ that is induced in a canonical way by
the geometry of $G$.

The space $\cF(G,\G, \Phi)$ of theories can be viewed as
stratified in terms of the group theoretic and automorphic
characteristics when $(G,\G, \Phi)$ are varied. This leads to
numerical data that identify the different strata, including
 the dimension and the rank of the group $G$, the type and level of the discrete group $\G$,
 as well as the weight of the automorphic forms $\Phi$. Depending on the specifics of the
 construction of $\Phi$, the automorphic structure provides additional data
 that characterize the field space.

Recent discussions in cosmology indicate that the basic structure
of automorphic field theory defines
  a natural framework in which to formulate inflationary models. It has been  emphasized
 in the literature that since inflation depends on small parameters
 constructed from the fundamental ingredients of the theory,
 phenomenological models are sensitive to the details of
 possible UV-completions, potentially invalidating their viability as realistic theories.
 An early discussion of these issues in the context of chaotic inflation
 \cite{l83} can be found in ref. \cite{em86}.
 A strategy adopted often in this context is to postulate the existence of an
 inflaton shift
 symmetry $\phi \longmapsto \phi+ s_\phi$ as a means to protect the potential from
 higher order
 corrections. The first model to incorporate this idea was natural inflation
 \cite{ni90, ni93}, but many
 modifications and extensions have been introduced in the intervening decades,
  including models
that aim at realizations of this symmetry in
 UV-complete theories \cite{s08etal, pp13rev, s13rev}.

In the context of the shift symmetry it is natural to ask whether this ad hoc operator
 is part of a symmetry group that acts on the inflaton field space.
 The existence of such a
group would provide a  systematic framework in which different
types of invariant potentials could be classified. A first step in
this direction was outlined briefly in \cite{rs14}, including a
concrete example of modular two-field inflation. It is the purpose
of the present paper to describe the general framework of
automorphic inflation
 as an automorphic field theory in more detail, including the discussion of
 classes of models
 with an arbitrary number of inflaton fields, leaving the phenomenological exploration of
 specific models to future publications. In Section 2 the general structure of
 automorphic
 field theories is  described, including the automorphic data that enters the
 construction of the
 inflationary potential, and the canonically constructed target space metric
 that is inherited from the symmetry group.  Section 3 contains a discussion of a
general class of automorphic forms given by
 Eisenstein series that can be used to construct automorphically invariant potentials.
 The case
 of the symplectic groups is used as an illustration. These groups have
 been of interest in the recent past in black hole physics and they provide a natural
 generalization of the modular group. Automorphic functions defined in terms of Eisenstein
 series illustrate how automorphic data provide further group theoretic information
 that helps to characterize the scalar field theory space.
 Section 4 specializes
  the framework of higher rank automorphic
 theories to the case of modular field theories, which provide an extensive
framework of two-field inflation. Section 5 discusses some of the
conclusions of this work, as well as some pointers to possible
extensions.

\vskip .2truein

\section{Automorphic inflation}

The general framework of automorphic inflation is given by multi-scalar field theories coupled to
 gravity and a nontrivial inflaton target space, characterized by a curved field space
 metric $G_{IJ}$
that is inherited from the symmetry group.

\subsection{Multi-field dynamics}

The action is given by the functional
 \beq
 \cA[\phi^I, G_{IJ}, g_{\mu\nu}]
  ~=~  \int d^4x \sqrt{-g} \left(\frac{M_\rmPl^2}{2} R
                                  ~-~  P(E^{IJ},V,\phi^I)\right),
 \lleq{mfi-action}
 where $M_\rmPl^2= 1/8\pi G$ is the reduced Planck mass,
  the signature of the spacetime metric is chosen to be
  $(-,+,+,+)$, and the metric $G_{IJ}$ on the configuration space $X$ of the scalar
  field multiplet $\phi^I$ will be discussed in more detail below.
 The function $P(E^{IJ},V,\phi^I)$ of
 $E^{IJ}~=~ \frac{1}{2}g^{\mu\nu} \del_\mu \phi^I \del_\nu \phi^J$
 and the potential $V$ is not necessarily rational.
 Discussions of the methods
 developed for the phenomenological analysis of multi-field
 inflation can be found in the references
  \cite{b05etal, w07, pt11, bhp12, k13etal, a14etal, b15etal, ww15, d15etal}
  and the literature cited therein.

The framework of automorphic inflation is general,
 but for many applications actions of the form
\beq
 P(E^{IJ}, G_{IJ}, \phi^I) ~=~ G_{IJ}E^{IJ} ~+~ V(F_i(\phi^I)),
 \eeq
 are of interest. Here the potential $V(F_i(\phi^I))$ is taken to 
 be a function of the inflaton multiplet $\phi^I$, determined
by functions $F_i$ on the bounded domain $X$ that defines the
target space of the coordinates $\phi^I$. The $F_i$
 are constructed from group-theoretic automorphic forms $\Phi_i$.

The metric $G_{IJ}$ on the target space $X$ is determined in a
canonical way from the group $G$ in a way that will be described
further below, leading to a Riemannian metric. The dynamics of the
system $(g_{\mu\nu}, G_{IJ}, \phi^I)$ involves the geometry of the
target space as well as that of spacetime via the Einstein
equations and the Klein-Gordon
 equation. Assuming that the covariant derivative on the inflaton space is of
 Levi-Civita type, the Euler-Lagrange form of the latter takes the form
 \beq
 \square_g \phi^I + \G^I_{JK} g^{\mu \nu}
 \del_\mu \phi^J \del_\nu \phi^K - G^{IJ} V_{,J} ~=~0,
 \eeq
where $\square_g = \frac{1}{\sqrt{-g}} \del_\mu \sqrt{-g} g^{\mu
\nu}\del_\nu$ is the d'Alembert operator and the $\G^I_{JK}$ are
the target space Christoffel symbols. This Klein-Gordon equation
can be written
 in terms of a covariant derivative $D_\mu$,   thought of as a combination of the
 spacetime Koszul connection
 \beq
 \nabla_\mu \del^\k \phi^I ~=~ \del_\mu \del^\k \phi^I + \G_{\mu \nu}^\k \del^\nu \phi^I,
 \eeq
 and a contribution of the curved target space
 \beq
 D_\mu \del^\k \phi^I ~=~ \nabla_\mu \del^\k \phi^I
    + (\del_\mu \phi^J) \G_{JK}^I \del^\k \phi^K
 \eeq
 as
 \beq
  D_\mu (\del^\mu \phi^I) ~-~ G^{IJ}V_{,J} ~=~ 0.
 \eeq

In a spatially flat Friedman-Lemaitre background the scalar field
dynamics takes the form
 \beq
  D_t \phidot^I ~+~ 3H\phidot^I ~+~ G^{IJ}V_{,J} ~=~ 0,
  \eeq
where  $D_t$ is the covariant derivative in target space given by
$D_t A^I = \del_t A^I + \G^I_{JK} \phidot^J A^K$. The Hubble
parameter is constrained in terms of the energy density
 $\rho=\frac{1}{2}G_{IJ}\phidot^I\phidot^J + V$
  via $H^2 = \rho/3M_\rmPl^2$ and
 $\Hdot = -G_{IJ}\phidot^I\phidot^J/2M_\rmPl^2$. Covariant
 formulations of
perturbations can be found in \cite{gt11, k13etal}, for example.

\subsection{Definition of automorphic field theories}

The notion of automorphic field theory derives from the idea to
consider dynamical systems whose structure is constrained by both
continuous and discrete symmetry groups in a systematic way.
 The original motivation to introduce such theories arose
 in the context of the shift symmetry, a discrete symmetry imposed in the inflationary
  context
in order to suppress higher dimension operator corrections of
phenomenological parameters, but the fundamental structure is
independent of such applications and the framework can be
formulated on its own terms, leading to the concept of
 automorphic field theories.

The classical concept of an automorphic function was originally
introduced in the late 19th century by Klein \cite{k1890} in the
context
 of the M\"obius group $G=\rmSL(2,\mathR)$ and its various discrete
  subgroups $\G \subset \rmSL(2,\mathZ)$.
  This framework is too narrow for general multi-field inflation,
   which motivates to consider extensions of this framework.
  In the early 20th century first generalizations
 by Hilbert, Siegel and Maa\ss~led to the introduction of generalized
 modular forms on higher-dimensional domains, as well as
 non-holomorphic forms, and since the 1950s work in this
 direction has experienced a vast
 generalization in the hands of Bochner, Selberg, Harish-Chandra,
  Langlands, and many others. As a result
 the concept has become less precise and different authors define it differently.
 Given the new conceptual picture developed for the general framework,
  it makes sense to distinguish between the classical case associated to the group
 $\rmSL(2,\mathR)$ (and $\rmGL(2,\mathR)$), which leads to
 two-field inflation, and the case of higher dimensional groups
 $G$ that are
 needed for true multi-field inflation. In the following the
 former case will be designated as modular, independently of the
 type of discrete groups $\G$ considered, and including the
 non-holomorphic forms of Maa\ss, and the case of higher
 dimensional groups $G$ will be called automorphic.

A natural class of groups that could be considered in this context
are semi-simple Lie groups,
 familiar from different fields in physics, but in order to include the general
 linear group
$\rmGL(n,\mathR)$, ubiquitous in the context of automorphic forms and
representation theory,
 it is useful to consider the more general framework of reductive groups.

\subsection{Automorphic potentials}

 The fundamental assumption in automorphic field theory is that the potentials
  $V(\phi^I)$ are invariant under some discrete symmetry group $\G$ acting on the
  scalar fields. In practice such functions are constructed in terms of group
  theoretic automorphic forms $\Phi_i$ and associated forms $F_i$ on the inflaton
  target space. These concepts will be made explicit in the following.

An automorphic form $\Phi$ is an object that here will be
 taken to be a scalar function
 \beq
 \Phi: G\lra \mathC
 \eeq
 on a reductive linear algebraic group, i.e. a group that can be embedded into
 the general linear
group $G \lra \rmGL(n)$.
  These functions are constructed in such a way as to admit a descent to the
  quotient space
 $
  X ~=~ G/K\cdot A
 $
 of the reductive group $G$ with respect to a maximal compact subgroup $K$ and the split
component $A$,  leading to a scalar form $F: X \lra \mathC$ on the
bounded domain $X$.

The inflaton multiplet $\phi^I$ is defined by spacetime functions
 that take values in $X$
 \beq
 \phi^I: ~M ~\lra ~ X,
 \lleq{scalar-fields}
 defining coordinates on the target space. A general inflationary potential $V$ can
 be viewed as
 a map on the configuration space $\cC^\infty(M,X)$ of all maps of the
 type (\ref{scalar-fields})
 \beq
 V:~ \cC^\infty(M,X) ~\lra ~ C^\infty(M,\mathR).
 \eeq
 In this set-up the potentials $V(\phi^I)$ of the multiplet inflaton $\phi^I$ are
 defined  as $V = V(F_i(\phi^I))$.

 Automorphic forms $F$ on the bounded domain
  $X$ can be constructed in terms of a special class of smooth group functions
  $\Phi$  on the algebraic reductive group with maximal compact subgroup $K$ relative to a
 discrete subgroup $\G \subset G(\mathZ)$. Different types of such discrete groups have
 been considered, in particular lattice subgroups, i.e. groups for which the quotient
  $\G\backslash G$ has finite volume, but it is most convenient to assume
  that $\G$ is an arithmetic subgroup, i.e. a discrete group that
  is commensurable to $G(\mathZ)$. If $\G$ is a subgroup of $G(\mathZ)$ then this implies
  that it has finite index in $G(\mathZ)$.
  If the group $G$ is semi-simple and of rank larger than unity
  then there is no distinction between lattices and
  arithmetic groups, but in general this is not case. The descent
  from the group $G$ to the domain is obtained by considering
  coset spaces of $G$ with respect to a maximal compact subgroup
  $K$, leading to a group theoretic input given by triplets $(G,\G, K)$.

  Automorphic forms associated to group triplets $(G,\G,K)$ are functions $\Phi$
  which can be characterized by a number of  constraints that are mostly concerned with the
 transformation behavior of $\Phi$. A further condition is usually formulated in terms
 of the universal
 enveloping algebra $U(\gfrak)$ associated to the Lie algebra $\gfrak$  by considering the
    quotient $T(\gfrak)/I(\gfrak)$  of the tensor algebra $T(\gfrak)$ of $\gfrak$ with
     respect to the
   two-sided ideal $I(\gfrak)$ generated by
  $V\otimes W -W\otimes V-[V,W]$, for $V,W \in \gfrak$.
 This constraint can alternatively  be viewed as a property of $\Phi$ encoded in terms of the
 algebra $D(G)$ of differential operators  because $U(\gfrak)$ and $D(G)$
   are isomorphic \cite{h78}.
 This follows via
 the iteration of the basic map that associates a differential operator
  $D_V$ on functions $\Phi \in C^\infty(G)$ to an element $V$ of the Lie algebra $\gfrak$
 \beq
  \gfrak ~\ni ~ V ~~\longmapsto ~~ D_V \Phi(g)
    ~:=~ \frac{d}{dt} \Phi(ge^{tV}){\Big |}_{t=0}.
 \lleq{la-vf-map}
 The most important part of the universal enveloping algebra is its center $Z_U$,
 for which the map
   (\ref{la-vf-map}) induces an isomorphism between $Z_U$ and
  the subspace of bi-invariant differential operators in $D(G)$.
  Briefly, an automorphic form on $G$ is then
  defined as a $\G-$covariant,  $Z_G$-covariant, $K$-finite, and $Z_U$-finite smooth
  function which
 satisfies a growth constraint. In more detail this entails the following.
\parskip=0pt
 \begin{enumerate}

 \item The first constraint encodes $\G$-covariance, the requirement that $\Phi$ transforms
     under a fixed arithmetic group $\G \subset G$ via a character $\e$ of $\G$, i.e.
 \beq
  \Phi(\g g) ~=~ \e(\g) \Phi(g), ~~~\g\in \G, ~g \in G.
 \eeq

 \item Relative to the center  $Z_G$ of the group $G$ the function $\Phi$ transforms in
 terms of
   the central  character $\om$
 \beq
  \Phi(ag ) ~=~ \om(a) \Phi(g), ~~~a \in Z, ~ g\in G.
 \eeq

 \item The requirement of $K$-finiteness demands that the space of functions spanned
       by transporting $\Phi$ via the right translation of the compact group $K$
       is finite dimensional
 \beq
 \rmdim ~\langle \Phi(gk)\rangle_{k \in K} ~< ~ \infty.
 \eeq

\item  The fourth constraint is concerned with the universal enveloping  algebra $U(\gfrak)$
    associated   to the Lie algebra $\gfrak$ of $G$.
     The requirement here  is  $Z_U$ finiteness,  which
    demands that the action of the center $Z_U$ of $U(\gfrak)$ generates a finite dimensional
    function subspace
  \beq
  \rmdim ~\langle D\Phi\rangle_{D\in Z_U} ~<~ \infty.
  \eeq
  Essentially this constraint generalizes the idea of considering eigenfunctions
  of the Laplacian of the group.

  \item The automorphic form satisfies a growth constraint.  There exists a norm
  $||\cdot||$ on $G$
  such that
 \beq
 |\Phi(g)| ~\leq ~ C||g||^n
  \eeq
 for all $g \in G$, where $C$ is a constant and $n\geq 0$ is an integer which
  depends on the choice of the norm $||\cdot ||$.
  If this is satisfied the function $\Phi$ is said to be of moderate growth.

  \item $\Phi$ is called a cusp form if
  \beq
  \int_{(\G\cap N)\backslash N} \Phi(ng) dn ~=~ 0
  \eeq
 for all $g\in G$ and all parabolic subgroups $P\subset G$ with Levi
 decomposition $P=MN$, where $N$ is the unipotent radical of $P$ and $M$ is called
  the Levi factor.
\end{enumerate}
More details can be found in ref. \cite{g58, hc59, hc68, bc79}.

 \parskip 0.15truein

Given a group function $\Phi$ of the type just outlined, it is
possible to construct the descent automorphic form $F$ on the
quotient $X=G/K$ which transforms under the discrete group in a
way that generalizes
 the behavior of classical modular forms. To do so one first notes that there exists
 a base point
 $z_0$ in $G/K$ such that the whole quotient is obtained via the action of the group
 $G$, i.e.
 $X=\{gz_0| g\in G\}$. The descent form $F$ can then be defined at any point
 $X\ni z=gz_0$ via
 \beq
  F(z) ~=~  J(g,z_0)\Phi(g),
 \eeq
 where $J(g,z_0)$ is a 1-cocycle function
 \beq
  J: ~G \times X ~\lra ~ \mathC
 \eeq
 which satisfies
 \beq
  J(g_1g_2, z) ~=~ J(g_1,g_2z) J(g_2,z).
\eeq
 The automorphy factor was introduced to generalize the older
notion of transformations determined by powers of the Jacobian
\cite{b51}.

\subsection{The kinetic term of automorphic inflation}

The class of automorphic potentials introduced above leads to  target spaces $X$ spanned
 by the scalar field multiplets $\phi^I$, $I=1,...,n$  that are homogeneous,
 i.e. the group of isometries acts transitively, i.e. any two points can be related
 by a group element. The metric $G_{IJ}$ on $X$  is taken to
be induced by the left-invariant metric of the group $G$.
 The left translation $L_h$ on $G$ for an arbitrary $h\in G$, defined by
 $L_hg = hg$, can be used to induce a metric on $G$ by using the differential
 $dL_g: T_eG \lra T_gG$ to transport tangent vectors at the identity $e$ of  $G$ to
 any other point $g$ of  $G$. It therefore can be used to extend any scalar
product
  $\langle \cdot, \cdot \rangle_e$ on $T_eG$ to the group
    via
 \beq
  \langle V,W\rangle_g ~=~ \langle dL_{g^{-1}}V, dL_{g^{-1}}W\rangle_e,
\eeq
 for $V,W \in T_gG$.  Since the tangent space $T_eG$ at the identity is isomorphic to the
 Lie algebra $\gfrak$ of $G$  via the map
 \beq
  V_g(f)~=~ \frac{d}{dt} f(ge^{tV}),
 \eeq
 where $V$ is an element in the Lie algebra  $\gfrak$, it is natural to
 define the scalar product on the tangent space $T_eG$ at the identity
 via the Cartan-Killing form $B(X,Y)$, defined in terms of the adjoint representation
             $\rmad_V: \gfrak \lra  \gfrak$ given by
             $W \longmapsto \rmad_V W:=[V,W]$, as
  \beq
 B(V,W) = \rmtr~(\rmad_V \rmad_W).
\eeq

The descent of the Cartan-Killing form induced metric on $G$ to the quotient $X=G/K$
can be
 constructed by considering the decomposition of the Lie algebra $\gfrak$ into the
 Lie algebra
 $\kfrak$ associated to the maximal compact subgroup $K\subset G$ and the
 orthogonal complement $\pfrak$
 \beq
  \gfrak ~=~ \pfrak ~\oplus ~ \kfrak,
\eeq
with respect to the form $B$.
The subalgebra $\pfrak$ is isomorphic to the tangent space at $z_0=eK=K$,
 and the restriction of $B$ to $\pfrak$ leads to an inner product on
 $\pfrak \cong T_{z_0}X$, which
can be transported with the left translation on the quotient
  space
    $\ell_h:  G/K \lra G/K$ given by $\ell_h gK =hgK$. This leads to
 \beq
 \langle U, V\rangle_z ~=~ \langle d\ell_{g^{-1}}U, d\ell_{g^{-1}}V\rangle_{z_0}.
\eeq

For matrix groups an explicit realization of the quotient space can be obtained
by mapping the
equivalence class $gK$ to $z=g(g^t)$ since $K$ is the kernel of this map. This
leads to a more
 immediate form of the scalar product as
 \beq
   \langle U,V\rangle_z ~=~ \rmtr(z^{-1}U z^{-1}V).
 \eeq
The metric associated to this scalar product is given by
 \beq
   ds^2 ~=~ \rmtr(z^{-1}dz z^{-1}dz).
\eeq

The local form $G_{IJ}$ of the metric on $X$ can now be constructed by
introducing coordinates
 on the quotient space that are inherited by the Iwawasa decomposition of the group $G$.
 This can be obtained from the Cartan decomposition of the Lie algebra
 $\gfrak = \kfrak \oplus \pfrak$
 by choosing a maximal  abelian subspace $\afrak$ of  $\pfrak$. The decomposition
 of $g\in G$
 then is induced by the exponential of the Lie algebra  decomposition
 \beq
 \gfrak ~=~ \nfrak \oplus \afrak \oplus \kfrak,
 \eeq
leading to $G=NAK$, where $N, A, K$ are the groups obtained from
$\nfrak, \afrak, \kfrak$ respectively.
 The quotient space is therefore spanned by $X\cong NA$ and the coordinates of $X$ can be
chosen to be those that appear in $N$ and $A$.

There are several classes of Lie groups that can be used to construct
automorphic field theories,
 including the semi-simple groups, but also reductive groups. A concrete
 sequence of groups
which illustrates the above abstract discussion are the symplectic
groups $\rmSp(n,\mathR)$.
 These groups most naturally generalize the model of $j$-inflation considered
 in \cite{rs14}
 as an example of modular inflation is given by the symplectic groups $\rmSp(n,\mathR)$,
 defined as the set of $(2n\times 2n)$-matrices $M$ such that
 \cite{m71, vdg08}
 \beq
  M^t J M ~=~ J, ~~~~~~J ~=~ \left(\matrix{ 0 &{\bf 1}_n\cr -{\bf 1}_n &0 \cr}\right),
 \eeq
 where ${\bf 1}_n$ is the $(n\times n)$-unit matrix and $M^t$ denotes the transpose of $M$.
 Writing $M$ in terms of $(n\times n)$-block matrices
\beq
  M ~=~ \left(\matrix{A &B \cr C &D\cr}\right) \in ~\rmSp(n,\mathR)
 \lleq{symplectic-matrix}
 translates the defining relation into the more transparent constraints
 \bea
   A^t D - C^tB &=& {\bf 1}_n \nn \\
   A^t C &=& C^t A \nn \\
  B^t D &=& D^tB.
 \eea
 For $n=1$ this specializes to the modular group $\rmSL(2,\mathR)$.
  The Iwasawa decomposition of $\rmSp(n,\mathR)$ is given by the compact subgroup
 $K= O(2n,\mathR) \cap \rmSp(n, \mathR)$, which is isomorphic to
\beq
  K_G~\cong~ \left\{\left(\matrix{A &B\cr -B &A \cr}\right) {\Big |} ~
          A^tA +B^tB ~=~{\bf 1},~~A^tB ~=~ B^tA\right\},
 \eeq
the block-diagonal subgroup
 \beq
  A _G= \left\{\left( \matrix{A &0\cr 0 &A^{-1} \cr}\right) ~{\Big |}~
   A~{\rm diagonal~ with~ positive~ elements} \right\},
  \eeq
  and
\beq
  N_G ~=~ \left\{\left(\matrix{A &B\cr 0 &(A^t)^{-1}\cr}\right)~{\Big |} ~
   A ~{\rm upper~triangular~with~1s~in~the~diagonal},~AB^t = BA^t\right\},
 \eeq
 where a subscript indicating the ambient group $G$ has been added in order
 to lessen the ambiguity of the notation. 
The compactness of $K$ can be seen by mapping $K$ to $U(n)$ via
 \beq
 \left(\matrix{A &B\cr -B &A \cr}\right) ~\longmapsto ~ A+iB.
 \eeq

The Cartan-Killing form leads to
 \beq
  B(V,W) ~=~ 2(n+1)  \rmtr VW
 \eeq
and the local form of the metric on the target space
$X=\rmSp(n)/U(n)$ then is given modulo a factor by
 \beq
  ds^2 ~=~ \rmtr~(C^{-1}dC)^2 + \rmtr~(C^{-1}dD)^2,
 \lleq{maass-metric}
  which  for $n=1$ specializes to the Poincar\'e metric of modular inflation discussed below.
 This metric
 is usually traced, without the group theoretic interpretation,
to Siegel's paper \cite{s43} but can be found in an earlier paper by
 Maa\ss~\cite{m42}.
 The map from $G$ to $X$ can be made explicit by choosing the maximal compact subgroup $K$
    to  be the isotropy group of a point $x_0 \in X$, leading to $gx_0 = nak x_0 = nax_0$.

Automorphic forms associated to the symplectic group have played an important role in the
 understanding of the microscopic structure of the entropy of certain types of
 supersymmetric black holes (refs. \cite{as08, rs13} contain reviews). Such forms can
 be obtained by lifting classical modular forms to Siegel forms and
 for the models considered so far the origin can in fact be
 traced to elliptic modular forms \cite{cs11}.

The same metric (\ref{maass-metric})
 is obtained for the target spaces $X=G/K$ with $G=\rmSL(n)$ and $K=\rmSO(n)$,
 where $A$ is given as above and
 \beq
  N ~=~ \left\{\left(\matrix{{\bf 1} &D \cr 0 &{\bf 1}\cr}\right)\right\},
\eeq
 where $D$ is symmetric.

\section{Automorphic functions}

As noted earlier, automorphic inflation is based on the idea to
consider potentials $V(\phi^I)$that are invariant with respect to
the action of a discrete subgroup $\G \subset G$. In principle one
might ask for a direct classification of the space of meromorphic
functions on the target space that are invariant with respect to
$\G$. A more practical and natural way to construct such functions
 is by considering quotients of automorphic forms of the same
weight with respect to $\G$. A special, but systematic, class of such
higher weight forms that can be used to construct quotients is comprised 
of  Eisenstein series associated to
 pairs $(G, \G)$, specified by further data.

\subsection{Eisenstein series}

In their classical incarnation Eisenstein series reach back into the 19th
century in the context of
elliptic geometry and modular forms associated to the modular group
$\rmSL(2,\mathZ)$.
 These objects are useful for two-field inflation,
providing in particular the ingredients for the potential used in
the model of $j$-inflation, an example of modular inflation
introduced in \cite{rs14}.
 The classical series  were generalized to the automorphic framework
 by several authors
 in the 1950s and 60s, in particular Selberg and Langlands,
 with a more detailed account published by Harish-Chandra \cite{hc68}.
 An extensive account of Eisenstein series in the adelic framework can
 be found in \cite{f15etal}. While the adelic language does have advantages
 for certain questions, and much of the modern literature on automorphic
 forms and representations is written in it \cite{rl76},
 this formulation is not always the best approach because it can obscure
 the geometric structures present in certain physical problems.
 The present description continues to use the archimedean framework to
 briefly outline the construction of
 Eisenstein series associated to higher rank groups, necessary for inflation
 with more than two fields.

 Roughly speaking, Eisenstein series are obtained by combining two basic constructions,
 the lift of functions defined on subgroups of $G$, and
 an averaging process of the resulting functions
  over quotients of discrete subgroups. The subgroups that enter the lift
  can be motivated as follows. If $G$ is a reductive algebraic group the
  Cartan decomposition allows to factorize the group as $G = PK$,
 where $K$ is a maximal compact subgroup. The remaining factor $P$ is not
 reductive, hence it contains a non-trivial unipotent radical $N$.
 Factoring further
    $P = MN$
 thus produces another, smaller, reductive group $M$,  called the Levi factor.
Defining a class of functions $I_s(G)$ by their transformation laws, which can be
made explicit more
easily in various special cases, the construction of $f_s$ depends in particular
on the nature of
the parabolic group $P$, which is encoded in the dimension of the parameter
$s\in \mathC^r$.
 For maximal parabolic groups $s\in \mathC$, while for minimal $P$ the
 dimension $r$ of $s$ is
given by the degree of the matrix group. Special cases can be
  found in Langlands \cite{rl89} and the modular case
 will be described in more detail later in this paper
 as a framework for two-field inflation.

The common feature of all cases is that certain "root" functions
 $f_M, f_K$ enter that are defined on the Levi component $M$ and the maximal compact
subgroup $K$, respectively
 \beq
  f_M: ~M \lra ~\mathC,~~~~~~f_K:~K ~\lra ~ \mathC.
 \eeq
These functions are extended trivially to $G$ and twisted by a character $\chi_s$
on the Levi component
 $\chi_s:M \lra \mathR$,
 which is also extended trivially. In the maximal case, i.e. for $s\in \mathC$,
 the character
 $\chi_s$ can be
 obtained as the $s^\rmth$ power of the canonically defined modular character,
 $\delta_P:  P \lra \mathR^+$,  which in this context is introduced to mediate between
 the left and
right Haar measures on $G$, i.e.  $d_Rg = \delta_P(g) d_Lg$.

Given the ingredients $(\chi_s, f_M, f_K)$, the lift of the functions on the smaller
groups, in particular
 the smaller reductive group given by the Levi component $M$,
 to the full group $G$ is obtained by defining
 \beq
 f_s(g) ~=~ \chi_s(m) f_M(m) f_K(k),
\eeq
 where $g= nmk$. The Eisenstein series $E_{s,f_P,f_K}$ are defined by averaging $f_s$
 over quotients
 of discrete groups where $P$ and $G$ are considered over the ring of integers $\mathZ$
 as
\beq
 E_{s,f_M,f_K}(g) ~=~ \sum_{\g \in P(\mathZ)\backslash G(\mathZ)} f_s(\g g).
\lleq{eisenstein-series}
  This construction of Eisenstein series
associated to reductive groups $G$ includes the holomorphic
 Eisenstein series  associated to the modular group $\rmSL(2,\mathR)$, as well as the
 real-analytic Eisenstein series  of $\rmSL(2,\mathR)$ introduced by
 Maa\ss.

The shift symmetry of the inflaton can be viewed as one of the two generators of the modular
group $\rmSL(2,\mathZ)$ and therefore leads to the consideration of modular inflation as a
framework for two-field inflationary models. The modular group can be embedded into higher
rank discrete groups $\G\subset G(\mathZ)$, such as subgroups of the general linear
 groups $\rmGL(n,\mathZ)$, or the symplectic groups $\rmSp(n,\mathZ)$, and thus allows
 an embedding of the shift symmetry in higher rank inflation. The resulting
 group structured spaces of scalar field theories are stratified further via
  the characteristics of automorphic forms, such as their rank,
weight, and level.

\subsection{Automorphic functions of higher rank}

As noted above in the geometric context of the kinetic term,
 the sequence of symplectic groups $\rmSp(n,\mathR)$ provides
 a natural extension of the modular group $\rmSL(2,\mathR)$,
 leading to automorphic inflation
based on Siegel modular forms.

 The symplectic group theoretic Eisenstein series were considered
 originally by Siegel \cite{s35, s39}.
 The domain $X$ in this case has dimension
 \beq
  \rmdim~\cH_n ~=~ \rmdim~\rmSp(n,\mathR)/U(n) ~=~ \frac{n}{2}(n+1),
  \eeq
 and is isomorphic to the Siegel upper halfplane defined as
 \beq
 \cH_n ~=~ \left\{Z=X+iY ~{\Big |}~ Z~{\rm symmetric}, ~Y>0\right\},
 \eeq
 where $Y>0$ means that $Y$ leads to a positve definite quadratic form. The
 symplectic group, with elements $M$ written as in (\ref{symplectic-matrix}),
 acts  on $Z \in \cH_n$ as
 \beq
 MZ ~:=~ (AZ+B)(CZ+D)^{-1}.
 \eeq
  The discrete group $\G$ can more generally be chosen to be a congruence subgroup
  $\G$ of the Siegel modular group $\rmSp(n,\mathZ)$, such as the Hecke
 congruence $\G_0^{(n)}(N)$ of level $N$, defined by the constraint that
 $C\equiv 0(\rmmod~N)$.
The dimension of the Siegel upper halfplane $X=\cH_n$ also
determines the number of algebraically independent functions on
$\cH_n$ that are invariant under the Siegel modular group
$\rmSp(n,\mathZ)$.

Generators of the space of automorphic functions with respect to the Siegel modular group
 of this space  can be constructed in terms of the series
 \beq
 E_{w,s}(Z) ~=~ \sum_{M\in \G_\infty^{(n)}\backslash \rmSp(n,\mathZ)}
      \frac{\rmdet~\rmIm(MZ)^s}{J(M,Z)^w},
 \eeq
which generalize the real-analytic Eisenstein series of genus 1.
 Here $w>n+1, Z \in \cH_n$, the symplectic automorphy factor
$J(g,z)$ is given by
 \beq
 J(M,Z) ~=~ \rmdet(CZ+D), ~~~~M = \left(\matrix{A &B\cr
 C&D\cr}\right) \in \rmSp(n,\mathR),
 \eeq
and
 \beq
  \G_\infty^{(n)}
   ~=~ \left\{\left(\matrix{A &B\cr 0 &D\cr} \right)\in
   \rmSp(n,\mathZ)\right\}.
  \eeq
  For $s=0$ these functions reduce to the series considered originally by
 Siegel \cite{s39}. Using the notation $E_w=E_{w,0}$ for these series one can consider
 the invariant functions
 \beq
 F_{rs}(Z) ~:=~ \frac{E_{rs}(Z)}{(E_s(Z))^r}
 \eeq
for $s>n+1$, which can be shown to generate modular Siegel modular
functions on $\cH_n$ \cite{s73, f83}.

It is also possible to consider more general Eisenstein series, following Klingen,
by lifting lower rank cusp forms. These are defined with respect to certain
 discrete subgroups  $\G_r^{(n)} \subset \G^{(n)}$,
 for $0\leq r \leq n$ \cite{k67, k90}, and cusp
 forms $F$ of degree $r$,
 evaluated at the projection $(M Z)_r$ of $M Z$ via
 \beq
  E_{w,r,F}(Z) ~=~ \sum_{M\in \G_r^{(n)}\backslash \G^{(n)}}
              ~\frac{F(M Z)_r}{\rmdet(CZ+D)^w}.
 \eeq
 It is therefore possible, for example, to
 lift classical cusp forms such as the Ramanujan form $\Delta = \eta^{24}(\tau)$
 (see below) to Siegel modular forms.

 The Siegel Eisenstein series $E_w(Z)$, $Z\in \cH_n$ specialize for $n=1$
 to the classical Eisenstein series of the upper halfplane. The latter were
 considered in the context of two-field inflation in \cite{rs14}.
  The next case in this sequence is given by genus 2
Eisenstein series for $\rmSp(2,\mathR)$, describing
 models of six-field inflation. The generators of this function field were first
 determined by Igusa \cite{i62, i64} in terms of Eisenstein series
 of weight 4 and 6, as well as three theta series of
 weight 10, 12 and 35, where the latter satisfies a polynomial
 equation in terms of the first 4 of these forms, and the weight 10 and 12 forms
 can be expressed in terms of the above Eisenstein series. This
 led in \cite{i62} to a set of generators of the function field involving forms
 of quite high weight. An alternative set of generators, involving lower
 weight forms was given by Freitag \cite{f65} in his simpler proof of
  Igusa's result, leading to
 \beq
  \frac{E_4E_6}{E_{10}}, ~~~~\frac{E_6^2}{E_{12}}, ~~~~ \frac{E_4^5}{E_{10}^2}.
 \eeq

The same structure can be extended to other groups, for example
 $\rmGL(n)$-automorphic inflation models based on $\rmGL(n,\mathZ)$-invariant
  functions that can be obtained from
  Eisenstein series associated to parabolic subgroups in $\rmGL(n,\mathR)$.

\vskip .3truein

\section{Inflation from modular field theory}

Over the past decade much effort has gone into the exploration of
a number of different two-field inflationary models with curved
target spaces, often with restrictions on the couplings of the
fields, such as sum-separable or product-separable potentials.
(See for example
  \cite{pt10, cap12, ht15, aaw15, bot15, gg15, eow15, kk15,  a15etal}
  and references therein.) The specialization of
automorphic field theory to two fields leads to new types of
classes of highly structured potentials involving classical
modular forms,
 leading to a stratified space of  two-field inflation models with curved
 target spaces. As noted earlier,
 classical modular forms were introduced in the second half of the
19$^\rmth$ century as functions
 on the complex upper halfplane because this space is mapped to itself by
 the modular group
 $\rmSL(2,\mathZ)$. Thinking about forms in this way is computationally useful,
 but conceptually limiting,
  and it is best to have both the domain function view and the group function view
  available.

 In the context of modular field theory the group $G=\rmSL(2,\mathR)$ is semisimple
 and the maximal compact  subgroup is the rotation group $K=\rmSO(2,\mathR)$,
 with the group action given by the M\"obius transformation. Thus the $(G,K)$
 structure is fixed
 for this class of models.  The  discrete groups $\G$ on the other hand can be chosen from
  various families that are  parametrized  by an integer $N$, the level. The theory
  has been developed
 in most detail for the Hecke congruence group $\G_0(N)$, defined by matrices
  $\g = {\tiny \left(\matrix{a &b\cr c&d\cr}\right)}$ in the modular group
  $\rmSL(2,\mathZ)$
such that $c\equiv 0(\rmmod~N)$. Such modular forms play an
important role in the context of geometric forms relevant for
motivic aspects of string theory \cite{rs06, rs08}.
 Other possibilities for $\G$
 include the groups $\G_1(N)$ defined by the further constraints $a,d \equiv 1(\rmmod~N)$,
  as well as the principal congruence subgroups $\G(N)$, which satisfy
  $\g \equiv {\bf 1}(\rmmod~N)$, i.e. $\G(N) \subset \G_1(N) \subseteq \G_0(N)$.
 More general classes of non-congruence groups can be considered as well.
 These will collectively be denoted as $\G_N$ in the following.
 The groups $\G_N$ thus characterize
 the field space of modular inflation in terms of their type $t_\G$ as well as their
 level $N=N_\G$.

\subsection{Modular potentials}

Modular forms on the group $\rmSL(2,\mathR)$ are functions of the type described
in Section 2.
 Given an arithmetic group $\G_N$ and a  character $\e_N$ the function $\Phi$ is required to
 transform as  $\Phi(\g g) = \e_N(\g) \Phi(g)$ for all $\g \in \G_N$ and
 $g\in \rmSL(2,\mathR)$.
The second constraint simplifies because for $k\in K$ and $g\in G$ one has
 \beq
 \Phi(gk) ~=~ \l(k) \Phi(g),
 \eeq
 where $\l_w$ is the weight character $\l_w(k) = e^{iw\theta}$ for
  $k={\tiny \left(\matrix{\cos \theta &\sin \theta\cr - \sin\theta &\cos\theta\cr}\right)}
    \in \rmSO(2,\mathR)$, where $w$ is the weight of the form.
  In particular, the $K$-translates span a 1-dimensional vector space.
 The behavior of the $G$-form with respect to the maximal compact subgroup
 thus introduces
 a further numerical characteristic that characterizes the scalar field theory space.
This automorphic part of the data is encoded more generally
 by a weight representation.

  The third constraint simplifies in the classical modular case because the action of
  the center of $\rmSL(2,\mathR)$ is trivial.
 The final condition simplifies as well
  because the center of the universal enveloping algebra
 is one-dimensional,  generated by the Casimir element $\cC=h^{ij} X_iX_j$,
 where $h^{ij}$ is the inverse of the metric
 defined on the Lie algebra $\gfrak$ via the Cartan-Killing form. Thus this constraint
 specializes to the requirement that the function $\Phi$ is an eigenfunction
 of the Laplace-Beltrami operator on $\rmSL(2,\mathR)$.
The eigenvalue structure distinguishes between holomorphic
classical modular forms and Maa\ss~forms. These constraints can be made explicit by
 considering the Iwasawa decomposition $G=NAK$ of $\rmSL(2,\mathR)$ in terms of the
 factors $N,A$ and $K$ defined above, which specialize here to
 \bea
   N &=& \left\{\left(\matrix{1 &x\cr 0&1\cr}\right)~{\Big |}~ x \in \mathR \right\}
         \nn \\
   A &=& \left\{\left(\matrix{\sqrt{y} &0 \cr 0&\sqrt{y}^{-1}\cr}\right) ~{\Big |}~
             y \in \mathR^+ \right\} \nn \\
   K &=& \rmSO(2,\mathR).
 \eea
The motivation for the specific form of the factor $A$ comes from fact that
the M\"obius action
 of $NA$ on the $K$-isotropy point $i$ gives $NAi = x+iy = z \in \cH$.

The Iwasawa decomposition in this case leads to
  $\rmSL(2,\mathR) \cong \mathR \times \mathR^+ \times [0,2\pi)$
 and the isotropy point $x_0=i$ establishes the isomorphism $NA \cong \cH$.
In these coordinates the Laplace-Beltrami operator reduces on the
  bounded domain $X=G/K$, which is isomorphic to the
 upper halfplane
 \beq
  X ~=~ \rmSL(2,\mathR)/\rmSO(2,\mathR) ~\cong ~
   \cH ~=~ \{\tau \in \mathC ~{\Big |}~ \rmIm(\tau) > 0\},
 \eeq
 to
 \beq
  \Delta ~=~ -y^2\left(\frac{\del^2}{\del x^2}+ \frac{\del^2}{\del y^2} \right),
 \eeq
where $\tau=x+iy$. Since the rank of $\rmSL(2,\mathR)$ is one, the center of the
universal enveloping
 algebra is spanned by the Laplacian,
  $Z_U(\gfrak) \cong \mathC(\Delta)$.

 Classical modular forms
  on $\cH$ are induced by group functions $\Phi: \rmSL(2,\mathR) \lra \mathC$
 via the  1-cocycle, given in this case by
 \beq
  J(g,\tau) ~=~ (c\tau + d), ~~~g=\left(\matrix{a&b\cr c&d\cr}\right).
 \lleq{sl2-cocycle}
 This leads to the definition
 \beq
   f(\tau) ~=~ (c\tau + d)^w \Phi(g),
 \eeq
 where $\tau = gi$. They are defined  with respect to discrete subgroups $\G_N$ of the
modular group $G(\mathZ) = \rmSL(2,\mathZ)$ and are characterized by the level $N$
 of the subgroup $\G_N$, their weight $w$, as well as a character $\e_N$ via their
transformation behavior, which for
 \beq
 \g ~=~ \left(\matrix{a &b\cr c &d\cr}\right) \in \G_N
 \eeq
 and the action of the M\"obius transformation
 \beq
 \g \tau ~=~ \frac{a\tau+b}{c\tau + d}
 \eeq
 leads to
 \beq
 f(\g \tau) ~=~ \e_N(d) (c\tau+d)^w f(\tau).
 \eeq

\subsection{Kinetic term}

The field space metric
 \beq
  ds^2 ~=~ G_{IJ} d\phi^I d\phi^J
 \eeq
 induced by the Cartan-Killing form $B$ on the Lie algebra
 $\sfrak\lfrak(2,\mathR)$ is the Poincar\'e metric, which in terms
 of the dimensionless variables $\tau^I=\phi^I/\mu$ via the energy
  scale $\mu$ can be written with $\tau = \tau^1+i\tau^2$ as
 \beq
  ds^2 ~=~ \frac{d\tau d\otau}{\rmIm(\tau)^2},
 \eeq
 defining a conformally flat target space geometry with
 \beq
  G_{IJ} ~=~ \left(\frac{\mu}{\phi^2}\right)^2 \delta_{IJ}.
 \eeq
 The kinetic term then takes the form
 \beq
  G_{IJ} g^{\mu \nu} \del_\mu \phi^I \del_\nu \phi^J
 ~=~ \frac{\mu^2}{\rmIm(\tau)^2} g^{\mu\nu}\del_\mu \tau \del_\nu
 \otau
\eeq and the action of modular inflation can be written as
 \beq
 \cA_\rmmod ~=~ \int d^4x\sqrt{-g} \left(\frac{M_\rmPl^2}{2} R
      - \frac{\mu^2}{2\rmIm(\tau)^2}
      g^{\mu \nu} \del_\mu \tau \del_\nu \otau
        - V(F_i(\tau))\right),
 \eeq
where the functions $F_i(\tau)$ are modular forms of weight $w_i$
on $\cH$.

\subsection{Geometry of the target space}

The Christoffel symbols of the Poincar\'e metric are given by
 $
 \G^1_{11} ~=~ \G^1_{22} ~=~ \G^2_{12}   ~=~ 0
 $
 and
 \beq
  \G^2_{11} ~=~ -\G^2_{22} ~=~ - \G^1_{12}  ~=~ \frac{1}{\mu \rmIm(\tau)}.
 \eeq

The bounded domain $X=\cH$ is two-dimensional, which implies via
the Bianchi identity that the Riemann curvature  tensor takes the
 form
 \beq
 R_{IJKL} ~=~ K(G_{IK}G_{JL} - G_{IL}G_{JK}),
 \eeq
 where $K=R/2$ is the Gaussian curvature expressed in terms of the Ricci scalar $R$.
 The Riemann tensor $R_{IJKL}$ has only one independent
 component
 \beq
  R_{1212}   ~=~ - \frac{1}{\mu^2} \frac{1}{(\rmIm~\tau)^4},
  \eeq
  leading to the constant curvature scalar $R=-2/\mu^2$.

\subsection{Modular Eisenstein series}

As in the automorphic case, Eisenstein series play a key role for modular forms of arbitrary
weight because for the full modular group they span the subspace complementary to the
 cusp forms. They also provide the building blocks for $j$-inflation \cite{rs14}.

 Holomorphic Eisenstein series are obtained by following the general construction
 briefly outlined in the general case in the previous section.
 The ingredients of the definition
 are obtained as follows. The group factorization  $G=PK$ into the parabolic factor and the
compact factor can be obtained from
 \beq
 P ~=~ NA ~=~
  \left\{\left(\matrix{\sqrt{y} &\frac{x}{\sqrt{y}} \cr 0 &\frac{1}{\sqrt{y}} \cr}
  \right) \right\}.
 \eeq

The function $f_K: K\lra \mathC$ is determined by the automorphy
factor $J(g,\tau)$ of eq. (\ref{sl2-cocycle}), restricted to the
 compact subgroup $K$ with base point $i=\sqrt{-1}$, and an integer $w$ as
 \beq
  f_w(k) ~:=~ \frac{1}{J(k,i)^w}.
 \eeq
The lift is characterized by a single complex parameter $s\in \mathC$ as
 \beq
  f_{s,w}(g) ~:=~ \delta_P(p)^{s/2} f_w(k),
 \eeq
 where the modulus character $\delta_P$ is defined for a
 $(2\times 2)$ matrix with the diagonal $(a,b)$ as
 \beq
  \delta_P(p) ~=~ \frac{a}{b}.
 \eeq

Denoting the discrete subgroup by $\G$ then leads to the Eisenstein series
associated to
 $\rmSL(2,\mathR)$
 \beq
  E_{s,w}(g) ~:=~ \sum_{\g \in (P\cap \G)\backslash \G} f_{s,w}(\g g)
  ~=~ \sum_{\g \in (P\cap \G)\backslash \G}
     \frac{\delta_P(p_\g)^{s/2}}{J(k_\g,i)^w}.
\eeq
This leads to both holomorphic and real-analytic Eisenstein series, in particular
the classical
 Eisenstein series
 \beq
  E_w(\tau) ~=~ 1 - \frac{2w}{B_w} \sum_{n\geq 1} \si_{w-1}(n) q^n,
 \eeq
 where $B_w$ are the Bernoulli numbers, $\si_w(n)$ is the divisor function
 \beq
 \si_w(n) ~:=~ \sum_{d|n} d^w
 \eeq
and $q=e^{2\pi i \tau}$. These functions generate the non-cuspidal part of the
space of modular forms.

\subsection{Modular functions}

Functions that are invariant with respect to subgroups
 $\G_N$ of the modular group $\rmSL(2,\mathZ)$ can be constructed by considering
 quotients of modular forms with respect
 to the same arithmetic group and the same weight. An
 important example
 is the quotient of two modular forms of weight twelve, the cube of $E_4$, and the
 Ramanujan modular form
 $\Delta$, which can be expressed in terms of the Eisenstein series as
 \beq
 \Delta(\tau) = \frac{E_4^3 - E_6^2}{1728}.
 \eeq
 Both forms are modular with respect to the full modular group
 $\G_1=\rmSL(2,\mathZ)$, but
contrary to the Eisenstein series, $\Delta$ is a cusp form.
 The latter appears in many different contexts, ranging from the bosonic partition
 function to the theory of motives.
 The quotient of these two weight 12 forms
  leads to the elliptic modular invariant
 \beq
  j(\tau) ~=~ \frac{E_4(\tau)^3}{\Delta(\tau)},
 \eeq
 an example of an $\rmSL(2,\mathZ)$-invariant function on $\cH$ constructed in a
 simple way from Eisenstein series. This function enters
  a number of  different applications, and an inflationary model based on
  $j(\tau)$ has been considered in \cite{rs14}.

Different types of modular forms  can be constructed in a
systematic way by considering modular forms for congruence
subgroups $\G_N\subset \rmSL(2,\mathZ)$. One extension is obtained
by defining Eisenstein series associated to such groups
\cite{m89}, while a second generalization involves the Dedekind
eta-function
 \beq
 \eta(q) ~=~ q^{1/24} \prod_{n\geq 1} (1-q^n),
 \eeq
 where $q=e^{2\pi i \tau}$. The latter can be used to consider
  products of the form
 \beq
  f(\tau) ~=~ \eta^{a_1}(q^{b_1}) \cdots \eta^{a_n}(q^{b_n})
 \eeq
 for appropriate constants $a_i, b_i \in \mathZ$.
 This also provides an alternative construction of the Ramanujan form as
 $\Delta(\tau)= \eta^{24}(\tau)$. Either type of congruence
 modular form can be used to construct modular functions via
 appropriate quotients.

\section{Conclusions}

Automorphic inflation can be seen as an application of a general
scalar field theory framework involving potentials constructed
from highly symmetric functions that are
 derived from automorphic forms associated to certain types of Lie groups $G$. The
 defining features
 include the choice of a discrete subgroup $\G \subset G$, examples of which are
 given by the
class of arithmetic groups. A further element of the construction
 is the choice of a maximal compact
subgroup $K$ of
 $G$, determined by a Cartan involution $\theta$. With these ingredients a special
 class of  functions
 is defined on $G$ whose transformation behavior induces automorphic forms on
 the quotient
 space $X=G/KA$, generalizing the usual transformation behavior via Jacobians.

The numerical characteristics of group pairs $(G,\G)$ include the
dimension $\rmdim ~G$ and the rank $\rmrk~G$ of the continuous
group $G$, as well as the level $N$ of the discrete group $\G_N$.
Further data is provided  by the weights $w_i$ of the automorphic
forms $(\Phi_i, F_i)$.
 One can therefore think of the space of  field theories
 as being stratified by the group structure $(G,\G_N)$, in which the different
 strata are characterized
 by a lattice of integers $(\rmdim~G, \rmrk ~G, ~N,~w_i)$.

A special class of inflationary theories that has received much
attention in past years is two-field inflation. Reducing
automorphic field theories to this context leads to modular field
theories, associated to
 the group of M\"obius transformation $\rmSL(2,\mathR)$ on the complex upper halfplane,
 leading to a framework for two-field inflation.  Here the dimension and rank of
 $G$ are fixed and
 the field theory space is stratified by the type of discrete subgroup $\G_N$,
 its level $N$, and the weights $w_i$ of the modular building
blocks.
  It would be interesting to explore the effect of higher levels in
  either the context of modular inflation models or in the general automorphic
  context, for example via symplectic Eisenstein series associated to
   congruence subgroups of the Siegel modular group \cite{b14}.

\vfill\eject

{\large {\bf Acknowledgements.}} \hfill \break
  It is a pleasure to thank Monika Lynker for discussions.
 This work was supported in part by the NSF grant 0969875.
 The author is grateful for the hospitality and support by the
 Simons Center for Geometry and Physics,
  where part of this work was done.

\vskip .4truein

\end{document}